# Hidden in Plain Sight For Too Long: Using Text Mining Techniques to Shine a Light on Workplace Sexism and Sexual Harassment[*]

[1]*Amir Karami*[†], [2]*Suzanne C. Swan*, [2]*Cynthia Nicole White*, [2]*Kayla Ford*
[1]**College of Information and Communications**
[2]**Department of Psychology**
*University of South Carolina*

**Abstract**

*Objective*: The goal of this study is to understand how people experience sexism and sexual harassment in the workplace by discovering themes in 2,362 experiences posted on the Everyday Sexism Project's website everydaysexism.com.

*Method*: This study used both quantitative and qualitative methods. The quantitative method was a computational framework to collect and analyze a large number of workplace sexual harassment experiences. The qualitative method was the analysis of the topics generated by a text mining method.

*Results*: Twenty-three topics were coded and then grouped into three overarching themes from the sex discrimination and sexual harassment literature. The *Sex Discrimination* theme included experiences in which women were treated unfavorably due to their sex, such as being passed over for promotion, denied opportunities, paid less than men, and ignored or talked over in meetings. The *Sex Discrimination and Gender harassment* theme included stories about sex discrimination and gender harassment, such as sexist hostility behaviors ranging from insults and jokes invoking misogynistic stereotypes to bullying behavior. The last theme, *Unwanted Sexual Attention*, contained stories describing sexual comments and behaviors used to degrade women. *Unwanted touching* was the highest weighted topic, indicating how common it was for website users to endure being touched, hugged or kissed, groped, and grabbed.

*Conclusions*: This study illustrates how researchers can use automatic processes to go beyond the limits of traditional research methods and investigate naturally occurring large scale datasets on the internet to achieve a better understanding of everyday workplace sexism experiences.

## Introduction

Workplace discrimination based on race, color, religion, sex, or national origin was banned by Title VII of the Civil Rights Act of 1964. While the number of sexual harassment charges filed to the Equal Employment Opportunity Commission has remained relatively stable over the last few years (Equal Employment Opportunity Commission, n.d.), recent reports of workplace sexual harassment with high profile people in Hollywood and politics have brought significant media attention to the issue. Researchers have found that while both men and women experience workplace sexual harassment, women are more likely than men to be targets of sexual harassment (Berdahl, 2007; Cortina, Magley, Williams, & Langhout, 2001), sexual

---

assault (Elliot et al., 2004), and gender-based discrimination (Schmitt et al., 2002; Eagly & Karau, 2002). There is no agreed upon national prevalence rate for sexual harassment, but researchers suggest that one out of every two women experience this form of harassment at some point in their working lives (Fitzgerald & Cortina, 2018). For example, McDonald's (2012) review found that roughly 13-31% of men and 40-75% of women experience workplace sexual harassment. While laws prohibiting sexual harassment have been implemented to stop this behavior in workplaces, it persists nevertheless.

The goal of this study is to further our understanding of how people experience and are affected by sexism and sexual harassment in the workplace by examining 2,362 workplace experiences posted on the Everyday Sexism Project's website everydaysexism.com. The Everyday Sexism Project was started by writer and activist Laura Bates in the United Kingdom in 2012 and "exists to catalog instances of sexism experienced on a day to day basis". Individuals posting on the website (referred to here as "users") enter their story into a text box; there is no word limit for the posts. If they choose, users can tag their entries with the categories workplace, public space, home, public transport, school, university, or media. Since the creation of the website, Bates has appeared in numerous media outlets around the world. A Google news search shows that this website has been considered by different news agencies such as the New York Times, Guardian, and The Economist. The approximately 100,000 posts on this website by Aug 2017 shows that a large number of people have voluntarily prompted themselves to discuss details of their stories.

To our knowledge, no other dataset containing information about sexual harassment is this large, diverse, and publicly available. Thus, we viewed the website as an excellent source of qualitative data to gather descriptive information about sexism and sexual harassment from a broad swath of people describing their experiences, including individuals who may not have the opportunity to share their experiences in a research study. We used a mixed method approach due to the large volume of data, as we explain in more detail below. First, we used a computational approach to identify categories of stories, based on a text mining method. Then, we applied a qualitative method, thematic analysis (Braun & Clarke, 2006), to ascertain the theme for each of the categories created by the computational approach, drawing from the sex discrimination and sexual harassment literature to code the data.

*Definition of sex discrimination and sexual harassment.* In the United States, the Equal Employment Opportunity Commission (EEOC) states that sex discrimination is a violation of Title VII of the Civil Rights Act, and is defined as "treating someone unfavorably because of that person's sex" (n.d.). Sex discrimination is a broad term that includes any unfavorable behavior in the workplace due to someone's sex, such as not hiring them, paying them less, giving them inferior work assignments, not promoting them, etc. Sexual harassment is a form of sex discrimination, defined by the EEOC (n.d.) as "Unwelcome sexual advances, requests for sexual favors, and other verbal or physical conduct of a sexual nature constitute sexual harassment when this conduct explicitly or implicitly affects an individual's employment, unreasonably interferes with an individual's work performance, or creates an intimidating, hostile, or offensive work environment." Some discriminatory workplace behaviors can be classified as sex discrimination, but not sexual harassment. An example is not inviting a female employee to important meetings and work trips that male employees are invited to. In the present paper, we use the term sex discrimination to refer to behavior that is discriminatory, but does not constitute sexual harassment.

*Types of sexual harassment.* In Fitzgerald & Cortina's (2018) comprehensive review of research on sexual harassment in the workplace, three broad categories of sexually harassing conduct have been delineated: *gender harassment, unwanted sexual attention*, and *sexual coercion*. As Fitzgerald and Cortina note, gender harassment expresses "insulting, degrading, or contemptuous attitudes about women" (p. 217). Gender harassment is not aimed at sexual cooperation; rather, the goal is to reinforce the inferior status of the gender being targeted (Leskinen & Cortina, 2014). Fitzgerald and Cortina describe four subcategories of gender harassment. *Sexist hostility* includes jokes, insults, and sexist comments. A second subcategory, *sexual hostility*, involves behaviors such as referring to women by degrading names of female body parts, displaying pornographic images, or crude comments about female sexuality (Stark, Chernyshenko, Lancaster, Drasgow, & Fitzgerald, 2002). A third subcategory, *gender policing*, expresses contempt for women who behave in ways not seen as stereotypically feminine. For example, gender policing was seen in the Price Waterhouse v. Hopkins Supreme Court case. The court found that Ann Hopkins (denied promotion despite performance reviews that were comparable or superior to her male colleagues) was subjected to sex discrimination when she was told to be considered for promotion, she needed to "walk more femininely, talk more femininely, dress more femininely, wear make-up, have her hair styled, and wear jewelry" (Price Waterhouse v. Hopkins, 1989, p. 235). A fourth subcategory, *work/family policing*, includes comments that women who have children should be in the home, not in the workplace, or that women with children are undependable employees and should not be trusted with important projects (Crosby, Williams, & Biernat, 2004). Gender harassment is the most widespread form of sexual harassment (Fitzgerald & Cortina, 2018; Leskinen, Cortina, & Kabat, 2011).

The second major category, *unwanted sexual attention*, refers to sexual advances that are uninvited and unwelcome. Such behaviors include asking for dates, comments about someone's body or attractiveness, attempts to establish a dating or sexual relationship with someone, unwanted touching, and sexual assault (Fitzgerald & Cortina, 2018). The final major category, *sexual coercion*, entails sexual advances in which the employee is offered some kind of benefit for acquiescing, or is threatened with a negative job-related consequence if they do not acquiesce (Fitzgerald & Cortina, 2018). For example, in a recent case at a Ford Motor Company plant, an employee who was repeatedly late for her 6 AM shift because her child's daycare center did not open early enough to allow her to get to work on time was told by her supervisor that if she had sex with him, he would not record the times she was late (Chira & Einhorn, 2017). These latter two categories differ in that unwanted sexual attention is not explicitly linked to a job benefit or consideration, whereas in sexual coercion a link to a job consideration for complying with the unwanted sexual request is implied or stated.

*Reporting sexual harassment.* Although sexual harassment is widespread, few victims file reports or official complaints (on average, less than 10%; Fitzgerald and Cortina, 2018). Often, reporting the harassment does not actually help, and may make things worse (Stockdale, 1998). Women who report sexual harassment can suffer both professional retaliation (work-specific actions such as undesirable reassignment or not being promoted) and social retaliation (being ostracized by coworkers or labeled as a troublemaker). Both types of retaliation are detrimental to professional and psychological well-being (Cortina & Magley, 2003). Despite legislation prohibiting retaliation for reporting harassment, it still occurs. For example, in a recent qualitative study on sexual harassment in the military, women reported being discouraged from reporting their experiences (Bonnes, 2017). One woman was told that she would lose her

Christmas leave if her superiors had to conduct an investigation and was asked if she really wanted to "ruin the man's career." The woman ultimately dropped the report.

*Use of communication technologies to study sexual harassment.* Because sexual harassment and sexism so often go unreported and hidden, it is easy to underestimate the scope of the problem. Recently, however, we have seen the impact of the Internet and social media on bringing sexual harassment to light (e.g., the #metoo movement). People's use of communication technologies, such as Twitter and online forums, have brought about a sea change in the public's awareness of the scope and reach of sexual harassment, providing an opportunity for victims to share their experiences. Importantly for our purposes, these technologies are also a new research opportunity to study large volumes of data related to sexual harassment and workplace sexism. Computational approaches address data collection and analysis barriers that often hinder traditional social science methods, allowing researchers to explore sexual harassment victims' electronic footprints at the societal level (Conte et al., 2012).

With massive datasets, it is impractical to analyze and organize large unstructured text data manually; therefore, text mining techniques are necessary to disclose hidden semantic features. Text mining can be defined as the methods of machine learning and statistics with the goal of recognizing patterns and disclosing the hidden information in text data (Hotho et al., 2005). Among text mining techniques, computational linguistics methods such as topic modeling (TM) are among effective and popular methods (S. Lee et al. 2010). TM is a statistical model for discovering topics in a corpus and represents the documents with those topics (Karami & Gangopadhyay, 2014; Karami et al., 2015b).

This study adds to the sexual harassment knowledge base by providing a framework to collect and analyze thousands of sexual harassment experiences from online resources such as everydaysexism.com. Other studies have utilized similar sites such as Twitter to get information on users' health opinions (i.e., Karami et al., 2018a; Fu et al., 2016; Scanfeld et al., 2010). Analyzing comments posted on websites allows researchers to go beyond the limits of traditional research methods (i.e., surveys, interviews, focus groups) and tap into people's descriptions of their experiences outside the context of a research study. This study draws from both data science and psychology approaches to mine the rich workplace sexual harassment data available on the Internet. We examined the following research questions: 1) Were the stories posted on the everyday sexism website consistent with the types of sex discrimination and sexual harassment identified in the research literature? 2) Were there any topics that emerged from the stories posted on everyday sexism that have not been identified in the research literature on sex discrimination and sexual harassment?

**Method**

*Participants.* Participants were users of the website www.everydaysexism.com who posted their experiences of workplace sexism and sexual harassment anonymously from March 2012 to August 2017. We selected the experiences having the "workplace" tag written in the English language and found 2,362 experiences[‡] in text format in this website. Regarding the adequacy of the sample size to examine our research questions, the 2,362 stories we used for our analysis is far greater than that used in other recent studies of sexual violence using comparable publicly available Internet data sources (e.g., Bogen, Bleiweiss, & Orchowski, 2019; Maas, McCauley, Bonomi, & Leija, 2018). In addition, our data collection approach doesn't have the risks of psychological harm inherent in asking questions aboutsensitive topics such as sexual harassment (Corbin & Morse, 2003). Avoiding these risks can also increase the data richness that is an important factor in selecting a proper sample size (Shrestha & Fawcett, 2016).

---

[‡] The collected data is available at https://github.com/amir-karami/WorkSpace_Sexual_Harresment

We do not know exactly how people who contributed their stories found out about the website, but we do know that the Everyday Sexism Project and its creator, Laura Bates, have received extensive media coverage. Her TED talk describing the project has over 1.1 million views, the Everyday Sexism Project Facebook page has over 31,000 followers, and the Everyday Sexism Twitter account has over 285,000 followers. Therefore we assume that website users found out about the website through social media or other media sources.

*Procedures*. We applied an advanced text mining approach, topic modeling, to disclose hidden topics in the collected data (a more detailed description of these methods can be found in the appendix). While different topic models have been developed, the latent Dirichlet allocation (LDA) is one of the most established and robust probabilistic topic models (S. Lee et al. 2010; Karami, 2015). This model clusters semantically related words in a text dataset. LDA assumes that each topic is a distribution over words and each document is a mixture of the topics in a corpus (Blei et al., 2003). LDA assigns the words that are semantically related to each other in a topic represented by a theme. For example, LDA assigns "job," "interview," "asked," "experience," and "role" to a topic that researchers could interpret as the "Job Interview" theme. This computational linguistic method has been applied in a wide range of applications such as analysis of health and medical corpora (Ghassemi et al., 2014; Paul and Dredze, 2014; Karami et al., 2018c), history data (Priva and Austerweil, 2015), spam detection (Karami & Zhou, 2014), business (Karami & Pendergraft, 2018), and Twitter data analysis in politics (Karami et al., 2018d; Karami and Elkouri, 2019), library (Collins & Karami, 2018; Karami & Collins, 2018), disaster management (Karami et al., 2019), and health (Shaw & Karami, 2017; Karami et al., 2018a; Karami et al., 2018b; Web et al., 2018; Shaw & Karami, 2019).

The outputs of LDA are the probability of each of the words (W) in each topic (T), $P(W|T)$, and the probability of each of the topics in each document (D), $P(T|D)$ (see Appendix). We also used $P(T|D)$ to find the total *sum of each topic*, $ST(T)$ and then *normalized* by the sum of the weight scores of all topics, $N\_ST(T)$ (see Appendix). The order of topics in the first column of Table 1 is based on the value of $N\_ST(T)$ from the highest (most mentioned topics) to the lowest (least mentioned topics).

In this study, we applied the Mallet implementation of LDA (McCallum, 2002), which was developed, based on Java programming language for text mining purposes. Mallet utilized preprocessing operations including tokenization, unigram representation, lower-casing, and removing numbers, punctuations, and the stopwords such as "the" and "a" that do not have semantic value for our analysis. We didn't remove any experiences and didn't use stemmers because they don't improve topic modeling significantly (Schofield & Mimno, 2016). To estimate the optimum number of topics, we used the ldatuning R package (Nikita, 2016) to compute $Log(P(W|T))$ (Griffiths & Steyvers, 2004) and found the optimum number of topics at 30 (see Appendix).

LDA is an unsupervised method and needs human judgment to interpret the overall theme in the topics. To conduct the qualitative analysis to determine the meaning of the topics, we used two sources of data. The first one is the list of the top 10 words for each topic (shown in Table 1). The second one is the top 20 stories for each topic based on $P(T|D)$.

*Analyses*. Once we had the words and top 20 stories for each of the topics, we proceeded to the final step, *topic analysis*, to ascertain the meaning of the topics. We used thematic analysis to conduct these analyses. Thematic analysis is a qualitative method for interpreting the conceptual meaning of themes (in this case, the topics; Boyatzis, 1998; Braun and Clarke, 2006). We used an inductive or 'bottom up' approach to analyze the themes, meaning "the themes…are strongly linked to the data themselves" (p. 83, Braun & Clarke), similar to that of Maas et al., (2018). To determine the theme for each topic, three authors separately coded each topic. The three coders then met to discuss the themes for each topic, using *consensus coding* (Lim, Valdez, & Lilly, 2015). Similar to Maas et al. (2018) and Lim et al. (2015), our consensus coding process was to meet, compare and contrast the themes we had generated separately, and keep on discussing them until we agreed on the final themes. For example, topic 16 contained these words: "*man car walking back street comment bus office lunch workplace*" (shown in Table 1). After each of the coders read the top 20 stories in this topic and coded it individually, we discussed it and reached consensus on coding this topic as "Commuting". See Table 2 for the labels and descriptions of each topic.

Table 1. Topics and Labels, Ranked by Weight (Highest to Lowest)

| Topic Number | Label | Topic |
|---|---|---|
| T1 | Unwanted Touching | back felt day hand put started night made touch kiss |
| T2 | Assumptions/Exclusionary | project boss team senior colleagues manager room meetings conversation office |
| T3 | Sexist Office | told looked office felt made point things conversation started nice |
| T4 | Gender discrimination | women sexism men workplace feel experience respect gender environment equality |
| T5 | Women are inferior | feel men don angry things women speak person point wrong |
| T6 | Reporting or telling | boss told workplace asked left things harassment police stop reported |
| T7 | Women's work | staff male female boss team manager job working role secretary |
| T8 | Normalizing sexism | male colleagues female work sexist fact response discussing make sexism |
| T9 | Sexualizing | comments made uncomfortable constantly talk joke body front part sexual |
| T10 | The customer is always | work manager shop man customers store day regular bar restaurant |
| T11 | Pay and opportunity | company manager position employees pay firm female senior ceo small |
| T12 | Job interviews | job interview asked experience role years worked engineering qualified department |
| T13 | Refusal to recognize | man work woman phone talking business asked idea times answer |
| T14 | Women's place | made good woman comments sexist point pretty coworkers kitchen joke |
| T15 | Service industry | told guys group drunk working pretty stop times girl afraid |
| T16 | Commuting | man car walking back street comment bus office lunch workplace |
| T17 | Sexual harassment and | harassment sexually behaviour inappropriate comments assaulted making body |
| T18 | Gender inequality | women men group successful business jobs gender equality expected status |
| T19 | Traditional family roles | children family mother home husband life good wife father parents |
| T20 | Maternity leave | work job leave woman weeks hard maternity days problem baby |
| T21 | Gendered honorifics | male called word doctor gender lady health hospital good nurse |
| T22 | Policing women's | dress hair short skirt makeup bra hot long size top |
| T23 | Sexist bullying | male big mine face show man open co-worker eyes laughed |

We then moved to the next step of the qualitative analysis, determining how many of the 30 topics should be retained. Again using the consensus coding method, we each separately identified topics that we thought should be deleted. Five topics were deleted because the stories in a topic were so different from each other that all three coders agreed that there was not a coherent theme for that topic. Two topics were deleted because the stories related to school and educational settings; for data reduction purposes, we decided to delete these and maintain the focus of the present paper on non-educational workplaces since most of the stories in our data took place in non-educational work settings. We met and discussed the topics to remove until we agreed on the final 23 topics.

In the final phase, coders used consensus coding and theoretical thematic analysis (Braun and Clarke, 2006) to group the 23 topics into themes. We coded the topics using the four types of sex discrimination and sexual harassment identified in the literature: 1) sex discrimination; and the three types of sexual harassment, consisting of 2) gender harassment (with the four types of gender harassment as subtypes (sexist hostility, sexual hostility, work/family policing, and gender policing); 3) unwanted sexual attention; and 4) sexual coercion. Table 2 shows the themes, subthemes, topic labels, and descriptions of each topic.

**Results**

As shown in Table 2, many topics contained stories that mapped onto more than one theme. The three overarching themes that can be used to characterize the topics are *Sex Discrimination*; *Sex Discrimination and Gender Harassment*, which contained topics that encompassed both of these constructs; and *Unwanted Sexual Attention*, which also contained several topics that included Gender Harassment. It is not surprising that topics often contained more than one type of sexual harassment, or contained both sex discrimination and sexual harassment, as other studies have found that different types of sexual harassment often co-occur (Leskinen, Cortina, & Kabat, 2011). To uphold the integrity of the experiences, grammar and spelling error were maintained.

***Sex Discrimination theme***. Six topics illustrated the theme of sex discrimination, as shown in Table 2. Sex discrimination includes sexist policies and practices, such as pay inequity, women being passed over for promotion by less-qualified men, and facing sexist comments in job interviews. For example, in *Assumptions/exclusionary practices*, women are not trusted to be able to perform tasks that are central to their jobs. One woman wrote, *"I provide a product recommendation to a customer and am asked to be transferred to someone who knows more about the product than I do. Male colleague provides exact same product recommendation and the customer is happy."* Other users wrote about being excluded from critical meetings and business trips: *"I work for a tech company where I am the only female on the 8 person leadership team…Recently I learned that my boss [invited] the other 7 male members of the leadership team …to join him on a trip to Las Vegas. I was not included on the invite…It was very awkward when this information came out and one of my colleagues said to me 'Do you want to grow a penis and join?'"*. This topic was weighted second out of the 23 topics, meaning it was the second most frequently mentioned topic in the data.

In *Women's work*, women are expected to perform traditionally female tasks in the workplace, even though it's not part of their job. The implication is that men are too important to perform these low-status activities: *"So just been promoted to the manager's team already been called a girl, asked to make tea when tidying, [and was] told it's 'woman's work.''* Another form of sex discrimination was shown in *Pay and opportunity inequity*. One woman's story provides a typical example of gender inequity: *"Passed over twice at work for a promotion in favor of a male colleague. Decided to undergo a work-based management qualification along with said newly promoted male colleague…My (male) Director told me that I 'need to have this qualification if I want to progress within the organization'. Well thanks. I guess it's either that or a penis right?"* In *Job interviews*, women described sex discrimination in interviews in which their competence and qualifications seemed to be overshadowed by their gender: *"After bein*

Table 2. Themes, Subthemes, Topics, and Descriptions of Topics

| Theme | Subtheme | Topic (Rank) | Description |
|---|---|---|---|
| **Sex Discrimination** | | Assumptions/Exclusionary Practices (2) | Assuming that women are not as competent or can't perform certain tasks. Downplaying women's accomplishments and not including women in important meetings. |
| | | Women's Work (7) | Women's work is being eye candy, being a secretary, tidying up, and making tea. Attempts to push women into these roles even when that's not their job. |
| | | Pay and Opportunity Inequity (11) | Paid less than a man, or not receiving a raise or opportunities that were originally promised. |
| | | Job Interviews (12) | Being treated differently in job interviews or being denied positions although more qualified than men. |
| | | Refusing to Recognize Women's Competence or Authority (13) | Inability to accept a woman's authority, asking to speak with a male colleague instead. Failure to acknowledge that a woman could be competent. |
| | | Policing Women's Appearance (22) | Women being told to make sure they aren't dressing too sexy for work because it would be distracting. Different clothing expectations for men and women. Being told to wear hair or makeup a certain way. |
| **Sex Discrimination & Gender Harassment** | **Sexist Hostility** | Gender Discrimination (4) | Sexism is so ingrained that both men and women subscribe to it. Women who bring it up are criticized for being too feminist or making something out of nothing. Assertive women are labeled as bossy or a bitch |
| | | Women are Inferior (5) | Women are seen as less competent than men. Women are talked over or not taken seriously. |
| | | Normalizing Sexism (8) | Men (and some women) do something sexist and then deny that the behavior is sexist. |
| | | Women's Place (14) | Coworker's comments about women's "proper place": domestic roles and for sex. Pet names: darling, honey |
| | | Gender Inequality (18) | Women are first fired, passed over for promotion, ignored at job fairs. No female speakers at the conference highlighting the "best and brightest" business executives. |
| | | Gendered Honorifics (21) | Denying women the honorifics they have earned (e.g., doctor) or using words that put women in their place (girl, babe, love) especially when they are competent (bossy, bitch). Refusing to accept women's authority. |
| | **Sexist Hostility & Sexual Hostility** | Sexist Bullying (23) | Men being condescending and demeaning to women. Objectifying women's bodies and calling them names. |
| | | Sexist Office Conversations (3) | Treating women like second-class citizens. Comments about women's bodies or sexuality. |
| | **Work/Family Policing** | Traditional Family Roles (19) | Women being pushed into traditional family roles, told they should focus on children instead of work. Men being criticized for taking responsibility for childcare. |
| | | Maternity Leave (20) | Experiencing discrimination due to pregnancy or motherhood. |
| **Unwanted Sexual Attention** | | Unwanted Touching (1) | Touching female colleagues without asking, or asking for contact that is not work appropriate. |
| | | Sexual Harassment and Assault (17) | Harassment and assault. Women expected to put up with it or change their behavior to avoid being assaulted. |
| | **Sexual Hostility** | Sexualizing (9) | Using sexual comments or behaviors to humiliate women in public workplaces. |
| | | The Customer is Always Right (10) | Customers making inappropriate comments to female employees and treating them differently than male employees. |
| | | Commuting (16) | Men whistling at, barking at, verbally harassing, catcalling, or making kissing noises at women who are commuting to or from work, or while out during lunch break. |
| | **Sex Discrimination** | Service Industry (15) | Directly harassed at jobs that require physical labor such as waitressing, bartending, carrying boxes/stocking/warehouse work. |
| | **Reporting Unwanted Sexual Attention** | Reporting or Telling Someone about Harassment (6) | Working up the courage to report harassment only for nothing to be done, or not reporting for fear of retaliation. |

*chosen...as the best candidate for a web developer job I was introduced to the CEO of the company. He asked me how is it possible that a woman becomes a developer since this job requires analytical skills."*

Another sex discrimination issue users wrote about is *Refusal to recognize women's competence or authority*: *"I traveled to India to do a site technical assessment ...[and] train my new Indian colleague... All the IT counterparts at the business are male. We introduced my colleague as a trainee at the start. I dive into my assessment questions. Every one of...my counterparts look away to my colleague and answer my questions to him. Or they ask clarifying questions to him. Takes about an hour and* Table 2. Themes, Subthemes, Topics, and Descriptions of Topics
*a half...before they finally figure out I am the one in the know and they finally start addressing me directly."*

In the *Policing Women's Appearance* topic, stories referred to the many complicated rules regarding women's attire. In some stories, women were warned about how they dress, move, and comport themselves, lest they "cause a distraction" or "appear unprofessional": *"In my late 40s and told that rumour has it a decade ago I was so hot that men in the office were distracted and didn't know where to look. Apparently my bra strap showed a lot and I wore pretty bras. Jeez over 10 years later and as an executive this what I am reduced to"*. In other stories, people felt they had the right to tell women how to dress or wear their hair: *"My ex-boss...noticed a few greys in my hair and demanded to know when I was going to dye them. A third of his own hair is silver but of course when women get greys dying them is priority #1"*.

**Sex Discrimination and Gender Harassment theme**. The second theme included topics that combined sex discrimination with the four forms of gender harassment. The first subtheme, *Sexist Hostility*, included both sex discrimination and sexist hostility. In the *Gender discrimination* topic, users wrote about double standards for men's and women's behavior: *"...It seems that it is acceptable for men to be assertive [and] their thoughts [can be] delivered in whichever way they choose but when women are 'assertive' in the same way...they are labeled aggressive..."* Gender discrimination was common, as this was the fourth-weighted topic in the data. Similarly, in *Women are inferior*, beliefs about women's inferiority were communicated in a variety of ways, including comments about women's driving *("A male colleague said recently that 'In his opinion' men are better drivers than women"*), women are crazy (*"...the only female manager of the business initiates ...a conversation about all women being crazy..."how much worse it was for men when women were in their periods"*), and women are unsuited for management positions (*"A male colleague saying to me 'It's nothing personal I just don't feel comfortable having a woman as a Manager."*). These types of comments were also frequent, as this was the fifth-weighted topic.

In the *Normalizing sexism* topic, users described how sexism is so normative and pervasive as to be invisible: *"I am a scientist...I post pictures on my social media about what I am currently...working on. This week I posted some electrical components and someone...asked me what my boyfriend was building. I responded and said it was mine and I also said that it was sexist for him to say it was my boyfriend's. To which he replied that he wasn't sexist. Then while telling my other male friend about the injustice he also told me it wasn't sexist ..."*. This topic also included sexist put-downs: *"Sitting at my desk with a colleague scrubbing the floor outside the office. Another (male) colleague came past and said jovially 'That's what I like to see: a woman on her knees!"*. Relatedly, *A woman's place* included coworker's comments about women's "proper place" (i.e., the kitchen or the bedroom). As this user wrote, *"As usual I'm the*

*only woman in the room. The instructor…was doing a bit of online shopping using the presentation screen, [he] came across a page with offers on kitchen supplies and felt compelled to say…'There you go Adrienne something for your kitchen!' As if I'm the only person in the room who could possibly be interested in kitchen supplies. Never mind that I'm the star of the class I really should get back to that kitchen.*"

In the *Gender inequality* category, users wrote about being passed over for promotion and ignored at job fairs. One user wrote: "*My workplace…has hired men from outside the company rather than let hardworking women from inside the company progress as it fits their profile and brand image better. I have seen these 3 men leave after a short period as they were not cut out for the job.*" This topic also included sexist hostility: "*A manager (man) at my workplace telling colleagues that women need to have children early in order to fit into men's schedules and if they don't…they're essentially worthless*". The last topic in the *Sexist Hostility* subtheme was *gendered honorifics*, in which women described being denied the honorifics they had earned. Many women wrote about these kinds of experiences in medical settings: "*I am wearing a stethoscope. I introduced myself as your Doctor and still you call me nurse*"! In addition to being denied titles, women wrote about being referred to in terms that undermine their professionalism: "*I don't go round calling men who are strangers 'babe' or 'love'…why do women get called these terms by men they don't know and men do not? Why does this happen in professional settings when I'm wearing a name badge*?"

The next subtheme under sex discrimination and gender harassment combined both *sexist hostility* and *sexual hostility*, as illustrated in the topic *Sexist bullying*. Some stories described sexual hostility:"*A young woman…who supplies art for the walls of the office came in to hang new pictures up. She had a drill and started holes in the wall. A very senior colleague of mine…leans back on his chair…stares at her…loudly declares how much he 'loves a woman with a power tool'…the poor woman had to keep doing her job…under the new horrible air of sleaze.*" Other stories described sexist hostility, bullying behavior that was sexist, but did not have sexual content:"*My…boss…asked me if I was doing my job correctly…I was testing magnets by holding them close to metal to see if the magnet held or came off…Not exactly a difficult task… Nevertheless, my boss stopped next to me…then tore open a completed box and began re-testing the magnets. He got bored quickly after finding that I had done it correctly and with a smirk insisted that I was 'lucky' to have done it correctly before walking away. I was left to clean up the mess he made in tearing the box open…He NEVER speaks so disrespectfully to my male co-workers.*"

The *Sexual Hostility* subtheme included the topic *Sexist office conversations*, sex discrimination behaviors that denied women privileges or opportunities due to their gender: "*In a new job…when on the Friday all of the men left for lunch at 1pm…one of these men had started in the office only a few days before me…Later told that I was not allowed to take lunch at 1pm on a Friday because of the standing Boys Pub Club! Ridiculed when I protested.*" This topic also included sexual hostility: "*In my previous job the office 'banter' was on a daily basis about the chest sizes of the women in the office or the way that we/female clients looked. When I…point[ed] out how sexist (and…rude) this was I was told (by both men and women) that I needed to 'lighten up'…Nobody else spoke out; some remained silent thus enabling the 'banter' to continue. I spoke to my manager and HR about this…Nothing was done.*" Notably, this was the third-highest weighted theme.

In the *Work/Family Policing* subtheme, users described how expectations regarding women's family roles impacted expectations of their work performance and career trajectories.

In *Traditional family roles*, users wrote about assumptions that, because they were women, having children would interfere with their work responsibilities: *"All of my colleagues are constantly asking me how I plan to balance having a family with my career and exams as a doctor. I have never mentioned wanting children."* In *Maternity leave*, many users wrote about the notion that once women have children, they can no longer be successful at their jobs: *"On being pregnant with my first child I was told by the COO of the company I worked for that I 'would be no bloodying use at all once that baby was born'"*. Pregnancy discrimination was also described: *"When my niece an engineer announced to her boss that she was expecting her first baby she found herself removed from her projects...A new male engineer was hired and given my niece's projects *and* a company car too (she doesn't have one)"*.

**Unwanted Sexual Attention theme**. This third theme included behaviors ranging from being asked for dates or drinks, to inappropriate sexual comments, to unwanted touching, to sexual assault. The *Unwanted touching* topic was the highest weighted type of workplace sexual harassment. One user wrote, *"... My boss at work is known to be a little creepy but today while talking to me in private he touched my thigh and ran it about a centimetre up my leg"*. In the topic *Sexual harassment and assault*, users described more extreme forms of unwanted touching and sexual assault: *"I was sexually assaulted by a man at my workplace...He grabbed my wrist and yanked me toward him causing my body to collide with his. He then used his other hand to grope my breasts and buttocks. He still had my right wrist in his grip and would not let go despite me fighting to get free...I shrieked right into his ear which startled him into letting go of me. I ran into an empty former patient room and locked the door so he could not get in...The next morning I complained to HR. He was not punished at all. He didn't get anything more than an instruction from HR to leave me alone"*.

In the *Sexual Hostility* subtheme of unwanted sexual attention, sexual comments and behaviors were used to humiliate women. In the *Sexualizing* topic, a user described how her coworker's behavior made her feel like an object: *"...[His] constant remarks about my appearance, how he would 'do me,' sexual innuendos, [and] invading my personal space. He even has pictures of me saved on his phone and tells me he 'likes to look at them?' It's hard to convey through writing how objectified this man makes me feel he is twice my age..."*. Another woman's story illustrates how her coworker humiliated her in front of their peers: *"I work in a bar....[coworker] came up behind me and pretended to do me doggy style from behind whilst holding my hair to get laughs out of the other boys"*.

*The customer is always right* refers to users being forced to endure customers who use the temporary power they have over them to harass and humiliate them: *"I was at work putting stock on low down shelves so...to not hurt my back by stooping all day I was on my knees. A regular customer came up to me and said "Oooh is that the position your boyfriend likes you in?"* In the *Commuting* topic, women posted about the unwanted sexual attention and sexist hostility they had to endure while commuting to work via public transit or walking. Often, women were treated as if they were animals to be whistled or barked at: *"On my lunch break walking across the road...get wolf whistled at. I ignore it and hear someone say 'bitch'...When I'm coming back from my lunch I consider walking a different...way in case they're still there but tell myself I'm being silly. Sure enough walking back across the road am wolf-whistled at again. And again when I just ignore it...I hear a man say 'stuck up bitch'."*

In the *Service industry* topic, users wrote about sex discrimination in the service industry, including comments that questioned their ability to perform physical labor that was required for their jobs: *"My boss wanted some tables moving...me and Greg are the same age and I'm the*

*one that goes to the gym. However my boss was insistent that the tables were too heavy for me to possibly lift and that we would just have to wait until Greg was free and that I should see if there's any washing up that needs doing…he seems to have forgotten that…I move all of those tables on a semi regular basis*". Users also wrote about the unwanted sexual attention that women in the service industry often have to endure: "*while waitressing at a corporate function last night a man asked how much he'd have to pay me to give him a lap dance*".

The last subtheme was *Reporting unwanted sexual attention*. The majority of users were reluctant to report harassment because of fear of retaliation or jeopardizing their jobs; responses to those who did report were typically not helpful. One woman wrote: "*When I was constantly having my ass slapped and groped by a guy at my work I finally worked up the courage to go to my manager…I told her I was too uncomfortable to confront him myself because he was higher up than me only to be told she couldn't do anything unless I spoke to him myself…This man is 20 years older than me and an aggressive person so I just let it happen until he…got bored and left me alone…*"

**Discussion**

While sexual harassment and sexism have long been a part of many women's (and some men's) working lives, it was rarely discussed publicly (McDonald, 2012; Fitzgerald & Cortina, 2018). Now, the Internet and social media are bringing sexual harassment into the open. Nowhere have we seen this phenomenon more dramatically than in the recent #metoo movement. Even before #metoo, the Internet's power to uncover hidden problems was discovered by thousands of individuals who visited the Everyday Sexism Project's website and posted their experiences of workplace sexism and sexual harassment. Our study uses a powerful and innovative text mining technique to analyze these experiences.

Addressing our first research question, the stories posted on the Everyday Sexism Project were largely consistent with the types of sex discrimination and sexual harassment identified in the research literature (EEOC, n.d.; Fitzgerald & Cortina, 2018). Drawing from this literature, we identified three overarching themes among these experiences: *Sex Discrimination*; *Sex Discrimination and Gender Harassment;* and *Unwanted Sexual Attention.* The *Sex Discrimination* theme included many experiences in which women were treated unfavorably due to their sex, such as being passed over for promotion, denied opportunities, paid less than men, ignored or talked over in meetings, and having their competence or authority questioned. Not surprisingly, stories about sex discrimination were often in the same topics as stories about gender harassment, as seen in the *Sex Discrimination and Gender Harassment* theme. Sexist hostility was prominent in this theme, represented in seven of the 10 topics. Sexist hostility behaviors ranged from various insulting comments and jokes invoking misogynistic stereotypes to challenging women's authority and refusing to refer to them by their titles, to bullying behavior. Illustrating the pervasiveness of sexism, the *normalizing sexism* topic contained stories of people doing something sexist and then denying that it is sexist, or not believing women's stories about the sexual harassment and discrimination they faced. In *Work/Family Policing*, women faced the presumption that, once they became mothers, they could no longer be successful in their jobs.

The last theme, *Unwanted Sexual Attention*, contained stories describing sexual comments and behaviors being used to degrade women. The *Unwanted touching* topic was the highest weighted topic, indicating how common it was for website users to endure being touched, hugged or kissed, groped, and grabbed. While users experienced unwanted sexual behavior in every imaginable profession, the service industry was frequently described and bears special mention. Service industry workers may be especially vulnerable to abuse and harassment, particularly from customers as they are often dependent on tips and customers are paying for not just the product, but the service as well. Particularly in bar and restaurant settings, when alcohol is involved, the ethos that "the customer is always right" can result in egregious abuse that workers have to endure as a condition of employment. One user's story illustrates this dilemma well: "*When I was a shot girl I was forever subjected to horrible comments and behaviour from not only the customers but even the bouncers of the club ...One night three bouncers pinned me against a wall and were laughing whilst they all pushed up against me they thought it was hysterical but I felt so degraded and helpless. Another night [a customer]..put his hand up my dress trying to finger me.. the job overall was horrible...but [I] felt slightly better because of the good money I was making it was a vicious circle*".

Sexual coercion did not emerge as a single topic in our data. In this form of sexual harassment, an employee is offered a benefit for acquiescing to a sexual advance, or is threatened if they do not. While sexual coercion stories did appear occasionally as a story in other topics, they were infrequent. This is consistent with the sexual harassment literature, which finds that sexual coercion is the least common type of harassment (Fitzgerald & Cortina, 2018).

Addressing our second research question, one topic that had not been identified in the research literature on sexual harassment that emerged from our data was the *Commuting* topic: unwanted sexual attention while commuting to or from work, on public transit or while walking (see Natarajan, 2016, for a recent study on this topic). While this type of harassment has been conceptualized as street harassment (Bailey, 2017), not workplace sexual harassment, in the sexual harassment literature, the website users in our data specifically chose the 'workplace' tag for their stories. To these users, getting whistled at, catcalled, or being touched or groped while commuting is *part of* the work experience. Before they even walked into their workplace, they have had to endure a gauntlet of sexual harassment.

Website users' stories about their decisions regarding reporting sexual harassment reflected how much victims have to lose, and how little they typically gain, from reporting. Over and over again, users who experienced sexual harassment wrote about remaining silent; this finding is highly consistent with the research on women's responses to harassment (Fitzgerald & Cortina, 2018; Stockdale, 1998). Although the sexual harassment made users feel violated and interfered with their ability to perform their job, complaining about the harassment was simply not seen as a viable option. They did not believe the risk of possibly losing their jobs, or getting even worse treatment for complaining, outweighed the unlikely possibility that something would actually be done to make the situation better. Indeed, the few users who did report sexual harassment to management or human resources typically stated that nothing was done, or that it did not help the situation.

**Limitations.** A limitation of the study is that we were not able to ascertain demographic information about users of the website, such as gender, age, or location. Thus, it is impossible to know very much about who the users are. When it was possible to ascertain the gender of users from cues in their stories such as names or pronouns, it appeared that the majority of users were women. This is not surprising given that the focus of Laura Bates' activism, as well as everydaysexism.com from which our data were collected, has been directed towards the sexist maltreatment of women. However, we know from previous studies that sexual harassment happens to men as well (e.g., Holland, Rabelo, Gustafson, Seabrook, & Cortina, 2016).

While we cannot know much about the demographic characteristics of the website users, we can speculate that many of them likely have some common characteristics, such as an interest in issues related to sexism and feminism, and a comfort level with using the Internet to express themselves. Thus, we make no claims that the experiences documented in the stories posted on the website represent all populations or workspaces. However, the experiences described in our study did map onto the types of sex discrimination and sexual harassment that have been documented in previous studies.

**Research Implications.** This interdisciplinary study is one example of how social scientists can study sexism and sexual harassment using computational techniques that allow for the analysis of naturally occurring, large scale datasets that are readily available on the internet (e.g., #metoo). Data such as this could be used to assess a variety of research questions, including the impact of these experiences on victims, workplace responses to victim complaints, and sexual harassment directed towards under-researched populations such as LGBTQ individuals along with considering other topic models such as structural topic models (Roberts et al., 2014).

**Policy Implications.** Consistent with previous studies, these data indicate that sexism and sexual harassment are widespread and pervade every aspect of many women's (and some men's) working lives. The results from this study could be used to develop training and improve existing workplace policies to better support victims. For example, in their guidelines for sexual harassment policies, the Society for Human Resource Management recommends that employers use examples of sexually harassing behavior, such as described in our results, rather than legal definitions (Segal, 2017). Additionally, this organization recommends that sexual harassment policies should extend to behavior by nonemployees, such as customers, also described in the results we present here. All in all, systemic interventions, such as fostering organizational climates in which everyone is respected (Fitzgerald & Cortina, 2018), and proactively addressing sexism and sexual harassment, hold promise as successful prevention strategies.

Acknowledgment
This work is partially supported by an Advanced Support for Innovative Research Excellence (ASPIRE) grant from the Office of the Vice President for Research at the University of South Carolina. All opinions, findings, conclusions and recommendations in this article are those of the authors and do not necessarily reflect the views of the funding agency.

# APPENDIX

**Topic Modeling**

Topic modeling is a text mining approach to discover hidden information in corpora (Aggarwal & Zhai, 2012). Among topic models. Latent Dirichlet Allocation (LDA) is the most popular topic model. This model shows a good performance to disclose the hidden semantic structure of a corpus (Karami et al., 2015a). LDA assumes that each of the documents (D) in a corpus has a mixture of topics (T) and each of the topics is a distribution of the corpus's words (W) (Blei, Ng, &Jordan, 2003). LDA converts the frequency of words in the documents represented by document-term frequency (DTM) matrix to two matrixes: (1) probability of words per each of the topics or P(W|T) and (2) probability of topics per each of the topics or P(T|D).

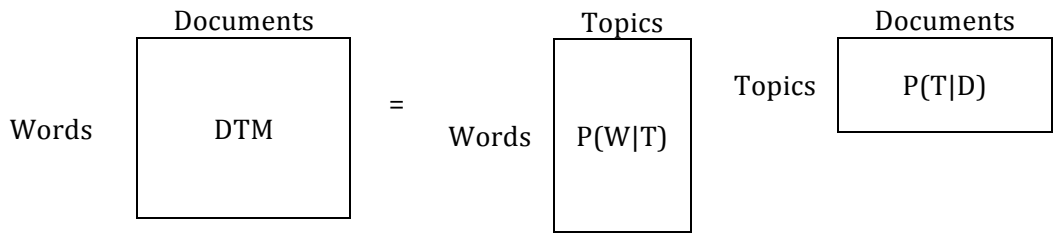

Figure A1: Matrix Factorization Interpretation of LDA

Each topic is a multinomial distribution of the words and each document is a multinomial distribution of the topics. For example in the following corpus with three topics, LDA assigns each of the words in the following document to each of the topics with a different degree of membership of probability. For instance in the following figure, "gene", "genetic", and "sequenced" were assigned to Topic 1 that can be interpreted as "Genetic" theme (Karami, 2015b).

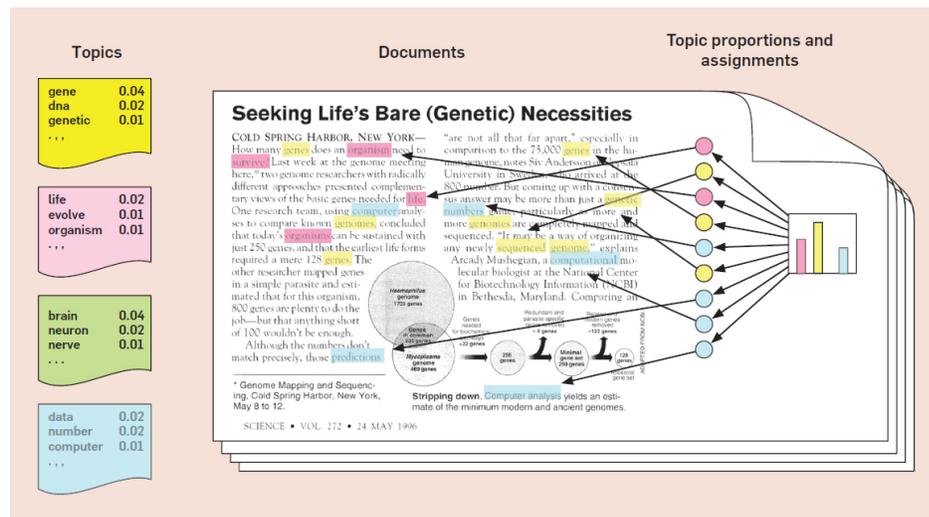

Figure A2: Intuition Behind LDA

The outputs of LDA for *n* documents (experiences), *m* words, and *t* topics are the probability of each of the words in each topic or $P(W_i|T_k)$ the probability of each of the topics in each document or or $P(T_k|D_j)$ (Karami, 2015).

$$\text{DTM} \rightarrow \text{Words} \begin{bmatrix} P(W_1|T_1) & \cdots & P(W_1|T_t) \\ \vdots & \ddots & \vdots \\ P(W_m|T_1) & \cdots & P(W_m|T_t) \end{bmatrix} \quad \& \quad \text{Topics} \begin{bmatrix} P(T_1|D_1) & \cdots & P(T_t|D_n) \\ \vdots & \ddots & \vdots \\ P(T_t|D_1) & \cdots & P(T_t|D_n) \end{bmatrix}$$

$$\text{Topics} \qquad\qquad\qquad\qquad\qquad \text{Documents}$$

$$P(W_i|T_k) \qquad\qquad\qquad\qquad\qquad P(T_k|D_j)$$

The top 10 words in each topic based on the order of $P(W_i|T_k)$ represent the topics. We also used $P(T_k|D_j)$ to find the total *sum of each topic* (S$T(T_k)$). To achieve an effective comparison, STs were *normalized* by the sum of the weight scores of all topics:

$$\text{N\_ST}(T_k) = \frac{\sum_{j=1}^{n} P(T_k|D_j)}{\sum_{k=1}^{t} \sum_{j=1}^{n} P(T_k|D_j)}$$

If $N\_ST(T_x) > N\_ST(T_y)$, it means that users shared more sexual harassment experiences about topic *x* than topic *y*. For extensive details and discussions on LDA, refer to Karami, 2015, Blei, 2012, and Steyvers & Griffiths, 2007. To find the top stories for each topic, we sorted $P(T_k|D_j)$ from the highest value to the lowest one. For example, we measured the $P(T_k|D_j)$ values for topic 1 and 2362 documents, $P(T_1|D_1),…, P(T_1|D_{2362})$, sorted them, and then found the top 20 documents (stories). It is worth mentioning that each topic has different membership values for each of the documents, indicating that each document can contain multiple topics. Different tools have been developed based on LDA in different platforms such as tm R package (https://cran.r-project.org/web/packages/tm/tm.pdf), lda python library (https://pypi.org/project/lda/), and Mallet java API (http://mallet.cs.umass.edu/topics.php). We utilized Mallet with its default setting for this paper. To apply Mallet on text data, a tutorial is available at https://programminghistorian.org/en/lessons/topic-modeling-and-mallet.

**Estimating Number of Topics**

To estimate the number of topics, different methods have been proposed. Four of them have been developed in ldatuning R package (https://cran.r-project.org/web/packages/ldatuning/vignettes/topics.html). In this paper, we utilized the log-likelihood approach (Griffiths & Steyvers, 2004) that is a common model selection method. Based on this approach, we computed log(P(w|T)) for T values of 10, 20, 30, 40, 50, 60, 70, 80, 90, and 100 topics. This process splits a corpus into train and test documents to estimate the probability of test documents given the train documents. The highest log-likelihood shows the optimum number of topics (Karami and Zhou, 2014). The result suggests the optimum point at 30 topics.

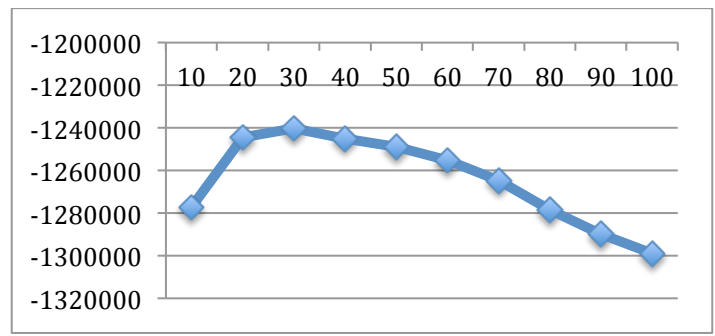

Figure A3: Estimating the Optimum Number of Topics